\documentstyle[psfig]{aa} 

\newcommand{\kms}{km~s$^{-1}~$}

\begin{document} 

\title{ A CO 
Survey of Gravitationally Lensed Quasars with the IRAM Interferometer 
\thanks{ Based on observations obtained with the Plateau de Bure 
Interferometer of the Institut de Radio Astronomie Millim\'etrique, supported 
by INSU/CNRS (France), MPG (Germany) and IGN (Spain). } } \author{ 
R. Barvainis\inst{1} D.
Alloin\inst{2} M. Bremer\inst{3} } \offprints{D. Alloin 
dalloin@eso.org} \institute{ National Science Foundation, 4201 Wilson Boulevard,
Arlington VA 22230, USA \and European Southern Observatory, Casilla 19001, 
Santiago 19, Chile \and IRAM, 300 Rue de la Piscine, 38406 St Martin 
d'H\`eres, France } 
\date{Received October 1, 2001, Accepted January 22, 2002 } \maketitle 
\markboth{R. Barvainis et al., 2002}{A CO Survey of Gravitationally Lensed 
Quasars}


{\bf Abstract.} We present the results of a CO survey of 
gravitationally lensed quasars, conducted with the Plateau de Bure 
Interferometer over the last three years. Among the 18 objects surveyed, one 
was detected in CO line emission, while six were detected in the continuum at 
3mm and three in the continuum at 1mm.  The low CO detection rate
may at least in part be due to uncertainties in the redshifts derived
from quasar broad emission lines.  
The detected CO source, the $z=3.2$ radio quiet quasar MG0751+2716,  
is quite strong in the CO(4$-$3) line and in the millimeter/submillimeter
continuum, the latter being emission from cool dust.  The integrated CO
line flux is $5.96\pm 0.45$ Jy$\cdot$kms$^{-1}$, and the total molecular 
gas mass is estimated to be in the range 
$M_{\rm H_2} = 1.6-3.1\times 10^9 M_{\sun}$.


\section{Introduction}

   The measurement of CO line emission in high redshift objects has
proven to be a fruitful avenue for investigating the
properties of distant quasars and galaxies. Molecular observations
address interesting issues such as that of star formation in the early
universe and how the presence of a central massive engine can affect
the interstellar material in its host galaxy. Detections of CO have
shown that quasars share some properties with luminous infrared
galaxies, both locally and at high redshift. For example, comparison
of the molecular, infrared and optical properties of the Cloverleaf
quasar and the infrared galaxy IRAS F10214+4724 (known to harbor
a buried quasar seen in polarized light) has demonstrated that these
two objects are nearly identical, except in the optical range where the 
differences can probably be attributed to obscuration/orientation effects 
(Barvainis et al.\ 1995). Such findings lend support to theories unifying 
luminous infrared galaxies and quasars via orientation effects, with high 
redshift infrared galaxies being the luminous counterparts of local Seyfert 
2's. However, on physical grounds, it also seems likely that IR-selected 
galaxies and UV-selected quasars may differ in their stage of evolution.  
There are currently about 15 well-documented detections of molecular gas at 
high redshift (e.g., Combes 2001), of which at least 9 are 
gravitationally lensed systems.
 
The advantages of using an intervening ``gravitational telescope'' to
boost the fluxes are obvious, with estimated magnification factors of
up to 100 in the optical. Moreover, differential gravitational effects 
provide an
elegant tool to probe the size and structure of the molecular material
within the quasar.  For example, a point-like emitting region (rest-frame UV
and optical continua from the inner accretion zone), and an extended 
dusty molecular region (the ``torus'') in the quasar will produce, after
gravitational effects from the intervening lens, images with different
morphologies.  Molecular line profiles, reflecting
intrinsic geometrical and kinematical properties,
can be particularly useful in understanding the extended structure. 
It should be noted however that a detailed model of the
intervening lens must be available to perform the transfer from the
image plane (observational data) to the source plane (intrinsic
properties of the quasar). We applied this technique for the
first time to recover the properties of the molecular torus in the
Cloverleaf, a quasar at $z=2.56$ (Kneib et al.\ 1998), comparing HST
images and IRAM interferometer CO maps (Alloin et al.\ 1997). The
CO-emitting region in the quasar was found to be a disk or ring-like
structure orbiting the central engine at a radius between 75 and 100
pc, with Keplerian velocity around 100 \kms . The effective resolution
resulting from this technique turned out in this case to be about 
20 times smaller than the synthesized beam size of the CO interferometer data.

Such significant benefits -- flux boosting and increased effective angular
resolution -- have led us to focus our attention on gravitationally
lensed systems and to conduct a CO survey of these objects. Another possible
benefit of lensed versus unlensed objects is the potential for higher flux 
boosting for higher-$J$ transitions (differential magnification).  Since
experience has shown that the best selection criterion for CO
detection is the presence of detectable far-IR or submm/mm dust
continuum emission, we started with a continuum survey of
the known lensed quasars using the IRAM 30m radio telescope and the JCMT
(Barvainis \& Ivison 2002). A high dust continuum detection rate encouraged
us to pursue a CO search with the IRAM interferometer 
(Guilloteau et al.\ 1992).  

Since this project was started early in the continuum survey, we
were not at that time able to make a general selection based upon submillimeter
flux.  Instead, the sample consisted of most of the then-known lensed
quasars having optical redshifts measured to good accuracy. 
We observed 18 gravitationally lensed quasars, with redshifts in the 
range $1.375-3.595$.

However,
reliable systemic redshifts remain a major difficulty for CO searches at high
$z$ because currently available redshifts are mostly derived from
highly ionized species in the quasar broad line region. As this region
is often coupled to a high velocity wind, redshifts derived this way 
have been found to be
blueshifted up to 1200~\kms with respect to the 
systemic velocity of
the host galaxy and the molecular environment of the quasar
probed by CO measurements.  A typical offset is  600 \kms , but there is 
wide dispersion from one object to another.  
Meanwhile, spectrometer bandwidths in the millimeter domain 
are too narrow ($\sim 1500$~\kms\ at 3mm) to fully span this 
redshift uncertainty 
using a single central frequency setting. The
combination of these two facts makes it likely that some CO 
lines will be missed in the course of a survey. In the case of the
present survey, whenever the quasar redshift was from highly
ionized species we applied a 600 \kms
redshift increment to search for its CO emission. We are fully aware that this
offset, although statistically meaningful, may be just incorrect for
some individual quasars.

  In Sect.\ 2 we describe the sample of gravitationally lensed quasars and
the acquisition and reduction of the
interferometer data set. Results, both in CO line emission and in
the 1mm and 3mm continua, are also presented in Sect.\ 2 for the entire 
sample. In
Sect.\ 3, we discuss the general results of the CO survey, and 
in Sect.\ 4 consider the detection of MG0751+2716 in the CO(4$-$3) 
transition in more detail. Conclusions and future prospects are 
given in Sect.\ 5.


\section{The sample: acquisition and reduction of the CO interferometric data 
set}

The gravitationally lensed quasar sample is presented in 
Table 1.   
The coordinates generally refer to the brightest quasar 
image in the optical. Redshifts have been corrected in some 
cases using the
technique described in the previous section, except for MG0751+2716
where an initial detection allowed refinement of the value to the center of the
line in followup observations.
The centering frequencies given for the 3mm
and 1mm windows correspond to various CO transitions from CO(2$-$1) to
CO(9$-$8), depending on the quasar redshift and the window considered:
the targetted CO transitions in the 3mm window are specified for each
source in Table 1.  
The ``seeing'' estimates correspond to the 3mm data set. 
Some targets were observed on several different observing runs;
seeing values, numbers of antennas, and hours spent per exposure are 
provided.  The observed spectral bandwidth was either 560 MHz or 595 MHz.

Table 2 lists the CO transition observed (3mm), the ``channel rms" 
(rms/beam/100~\kms , in 
mJy), the line flux (for MG0751+2716), and the 3mm and 1mm continuum
results.   Sufficient sensitivity and
bandwidth for line measurements were only available at 3mm.  
Table 2 also lists 850~$\mu$m continuum fluxes or upper limits derived 
from SCUBA observations at the JCMT (Barvainis 
\& Ivison 2002).
Note that B1600+434 and B1030+074 are strong radio loud quasars whose 
millimeter/submillimeter continua are consistent with pure synchrotron 
radiation. The other 
submillimeter-detected sources are likely to be dominated by dust emission. 
The continuum flux ratio between 3mm and 1mm for MG0751+2716 is not 
typical of dust emission, though the ratio between 1mm and 850~$\mu$m is.
This is probably caused by some residual steep-spectrum synchrotron emission
contributing to the 3mm (and to a lesser extent the 1mm) flux. 

\begin{table*}
\label{tab-sample} 
\caption{Target list and observing parameters of the CO survey} 
{\tiny
\begin{tabular}{l|llcrlllllll} 
Object & RA(2000) & 
Dec(2000) & z & $\nu_{\rm 3mm}$ & Seeing & Beam size/PA & $\nu_{\rm 1mm}$ 
& Beam size/PA & Antennas & Exposure\\ 
\hline & & & & (GHz) & ($\arcsec$) & ($\arcsec\times\arcsec/^{\circ}$)
& (GHz) & 
($\arcsec\times\arcsec/^{\circ}$) & & (Hrs)\\
\hline 
0047-2808 & 00:49:41.87 & $-$27:52:25.7 & 3.595 & 100.335 & 1.6;1.0 & 
$14.1\times 3.5/11$ &  225.661 & $-$ & 4;5 & 3.5;1.6 \\ 
UM673 & 01:45:17.22 & $-$09:45:12.3 & 2.730 & 92.706 & 0.75 & 
$4.5\times 1.9/0$ & 216.260 & $1.9 \times 0.7/199$ & 5 & 4.2 \\ 
MG0751+2716 & 07:51:41.46 & $+$27:16:31.4 & 3.200 & 109.778 & 1.0;1.1;0.6& 
$9.4\times 5.6/32$ & 246.898 & $3.1\times 2.1/14$ & 4;3;4 & 3.4;3.6;2.0 \\ 
SBS0909+523 & 09:13:00.76 & $+$52:59:31.5 & 1.375 & 97.068 & 0.2 & 
$2.4\times 1.7/155$ & 242.639 & $0.9\times 0.7/147$& 5 & 4.7 \\ 
RXJ0911+055 & 09:11:27.50 & $+$05:50:52.0 & 2.807 & 90.831 & 0.7 & 
$6.5\times 6.0/53$ & 211.886 & $3.0\times2.0/127$& 4 & 6.3 \\ 
Q1009-0252 & 10:12:16.09 & $-$03:07:03.0 & 2.746 & 92.311 & 0.7 & 
$13.0\times 6.2/9$ &  215.337 & $5.5\times 2.9/15$& 4 & 3.1 \\ 
J0313 & 10:17:24.13 & $-$20:47:00.4 & 2.552 & 97.352 & 0.9 & 
$17.4 \times 6.3/7$&  227.097 & $8.0\times2.4/7$ & 4 & 4.0 \\ 
B1030+074 & 10:33:34.0 & $+$07:11:26.1 & 1.535 & 90.942 & 1.1;2.3 & 
$9.9\times 5.8/49$ &  227.325 & $-$ & 5;5 & 1.1;3.2 \\ 
HE1104-1805 & 11:06:33.45 & $-$18:21:24.2 & 2.326 & 103.967 & 1.7 & 
$13.5\times 4.5/8$ &  242.529 & $-$ & 3 & 3.2 \\ 
PG1115+080 & 11:18:16.96 & $+$07:45:59.3 & 1.723 & 84.663 & 0.7 & 
$7.8 \times 5.5/53$ & 211.630 & $2.8 \times 2.2/60$ & 4 & 5.8 \\ 
1208+1011 & 12:10:57.16 & $+$09:54:25.6 & 3.831 & 95.434 & 
0.5 & $2.5 \times 1.4/23$ & 214.637 & $1.1\times 0.6/24 $ & 4 & 6.3 \\ 
HST14176+530 & 14:17:36.51 & $+$52:26:40.4 & 3.403 & 104.711 & 
0.8;2.0 & $6.6\times 4.8/50$ &  235.501 & $2.4\times 1.7/98$ & 4;5 & 3.7;5.9 \\ 
SBS1520+530 & 15:21:44.83 & $+$52:54:48.6 & 1.860 & 80.608 & 
1.1;0.7;0.7;1.2 & $10\times 5.7/123$ & 241.773 & 
$5.3\times 2.5/63$ & 4;3;5;5 &  4.4;3.8;0.33;2.3 \\ 
B1600+434 & 16:01:40.45 & $+$43:16:47.8 & 1.589 & 89.045 & 
0.2;1.4;2.0 & $8.3\times 7.6/138$ & 222.583 & 
$1.0\times 0.6/46$& 5;4;4 & 1.9;6.1;2.9 \\ 
B1608+656 & 16:09:13.96 & $+$65:32:29.0 & 1.394 & 96.298 & 0.4 & 
$3.0\times 1.7/83$ &  240.713 & $1.2\times 0.7/75$ & 5 &  3.9 \\ 
2016+112 & 20:19:18.15 & $+$11:27:08.3 & 3.282 & 107.670 & 0.7;0.6;1.3 & 
$7.8\times 5.0/45$ &  242.156 & $3.0\times2.4/21$ & 5;5;4 & 4.9;2.6;4.9 \\ 
HE2149-2745 & 21:52:07.44 & $-$27:31:50.2 & 2.033 & 114.011 & 
3.6 & $24.1 \times 4.6/7$ & 227.938 & $-$ & 5 & 3.0 \\ 
Q2237+0305 & 22:40:30.14 & +03:21:31.0 & 1.696 & 85.511 & 1.0 & 
$6.5 \times 4.1 /2$ 
& 213.749 & $2.6\times 1.6/176$ & 4 & 5.1 \\ 
\end{tabular} 
}
\end{table*}

\begin{table*}

\caption{Results of the CO survey} 
\label{tab-results} 
\begin{tabular}{l|lccllll} 
Object & Target line & Line rms & Line flux & 3mm $S_{\rm cont}$ &
1mm $S_{\rm cont}$ & $850\mu$m $S_{\rm cont}$ \\ 
\hline 
 & (mJy/beam)  & (mJy/100kms$^{-1})$& (Jy$\cdot$kms$^{-1}$) & (mJy/beam)& (mJy/beam)
& (mJy) \\
\hline 
0047-2808    & CO(4$-$3) &3.3& $-$ &0.9 $\pm$ 0.9 & na &$<7.0$ \\ 
UM673        & CO(3$-$2) &2.0& $-$ & 0.7 $\pm$ 0.5 & na  &$12.0\pm 2.2$ \\ 
MG0751+2716  & CO(4$-$3) &1.9& $5.96\pm0.45$ & 4.1 $\pm$ 0.5 &6.7 $\pm$ 1.3 &$25.8\pm 1.3$ \\ 
SBS0909+523  & CO(2$-$1) &1.1& $-$ &0.5 $\pm$ 0.3 & 0.0 $\pm$ 1.3 &$<5.5$ \\ 
RXJ0911+055  & CO(3$-$2) &1.2& $-$ &1.7 $\pm$ 0.3  & 10.2 $\pm$ 1.8 & $26.7\pm 1.4$ \\ 
Q1009-0252   & CO(3$-$2) &1.8& $-$ & 0.1 $\pm$ 0.5 & 0.3 $\pm$ 4.0 &na \\ 
J0313        & CO(3$-$2) &1.8& $-$ &0.0 $\pm$ 0.5 & -6 $\pm$ 4 &na\\ 
B1030+074    & CO(2$-$1) &3.3& $-$ & 184 $\pm$ 1.9 & na &na \\ 
HE1104-1805  & CO(3$-$2) &6.8& $-$ & 1.9 $\pm$ 2.0 &na &$14.8\pm 3.4$\\ 
PG1115+080   & CO(2$-$1) &1.3& $-$ & -0.3 $\pm$ 0.3 & -0.9 $\pm$ 1.2 &$3.7\pm 1.3$ \\ 
1208+1011    & CO(4$-$3) &1.3& $-$ &0.2 $\pm$ 0.3 & 4.2 $\pm$ 1.9 &$8.1\pm 2.0$\\ 
HST14176+530 & CO(4$-$3) &1.2& $-$ & 0.1 $\pm$ 0.3 & -5 $\pm$ 3 & $<3.5$\\ 
SBS1520+530  & CO(2$-$1) &1.5& hint &  0.0 $\pm$ 0.2 &-0.5 $\pm$ 1.4 &$9.4\pm 2.6$\\ 
B1600+434    & CO(2$-$1) &1.1& hint &25 $\pm$ 0.3 & 12.6 $\pm$ 2.3 &$7.3\pm 1.8$ \\ 
B1608+656    & CO(2$-$1) &1.5& hint & 8.1 $\pm$ 0.4 &0.8 $\pm$ 2.2 &$8.1\pm 1.7$\\ 
2016+112     & CO(4$-$3) &0.8& $-$ &1.8 $\pm$ 0.2 & 1.1 $\pm$ 1.0 &$<4.8$\\ 
HE2149-2745  & CO(3$-$2) &10.& $-$ &-2.8 $\pm$ 2.7 & na &$8.0\pm 1.9$\\ 
Q2237+0305   & CO(2$-$1) &1.4& $-$ &0.1 $\pm$ 0.35 & -1.9 $\pm$ 2.1 &$3.9\pm 1.2$ \\ 
\end{tabular} 
\end{table*}

All of the continuum detections, and the one line detection, are within 
$\sim 1\arcsec$
of the optical positions listed in Table 1, except for two cases.  
For 2016+112, the offset is $\delta\rm{RA}
=-1\farcs5$ and $\delta\rm{Dec}=3\farcs5$ from the observed coordinates.
However, checking the NASA Extragalactic Database (NED) we found 
improved coordinates which are within about $1\arcsec$ of the
continuum source in both RA and Dec.
In the case of RXJ0911+055, the offset of the continuum
source (a mean of the 3mm and 1mm continuum sources) is $\delta\rm{RA}
=2\farcs0$ and $\delta\rm{Dec} =2\farcs1$.  The origin of this 
offset, which is larger than typical optical position errors, is 
unknown at present.

\section{Results of the CO survey}

Out of 18 sources
observed, we obtained: (a) one strong CO line emitter, MG0751+2716, detected
in CO(4$-$3), (b) three marginal detections (to be
investigated further) and (c) 14 non-detections. 
Generally speaking, strong submm emission from dust is 
a good predictor of strong CO emission.  Of the two strongest dust 
sources, RXJ0911+055 ($S_{850} = 26.7$ mJy) and MG0751+2716 
($S_{850} = 25.8$ mJy), only the latter was detected in CO.
The line, with a 
peak flux of 18 mJy, is one of the strongest known among high-$z$ sources.  
Seven sources had either no measurements or only upper limits at 
850~$\mu$m, but 6 additional sources had moderately strong submm 
continuum detections in the $8-15$ mJy range.  

We believe some CO lines, such as the strong
one expected from RXJ0911+055, and some among the intermediate-strength 
submm sources, may have been missed 
because of the uncertainty in the redshift of the molecular 
lines combined with the narrow observing bandwidth.  In fact, during our
1999 first search for CO in MG0751+2716, the CO(4$-$3) transition was
detected on the edge of the bandpass, at a velocity offset of $-600$
\kms with respect to our original guess for molecular emission. Thanks
to the strength of this line we were able to identify its presence and
obtain new observations at the appropriate frequency, which fully
confirmed the 1999 measurement. However, this would not work 
in the case of fainter CO
line emitters. This difficulty will only
be fully resolved by using broadband backends in the future. In the 
meantime, we plan to reobserve at least RXJ0911+055 with flanking 
bandpasses to cover more redshift space. 

Continuum emission was detected at 3mm in 6 targets, with 3 of those
also detected at 1mm (one being particularly strong, the radio loud 
quasar B1030+074). 

\section{The CO line in  MG0751+2716}

A strong line in the CO(4$-$3) transition was detected in
MG0751+2761. A set of the channel maps, with velocity steps of 100
km s$^{-1}$, is shown in Fig.\ 1 for the CO(4$-$3) transition. It 
convincingly reveals the CO(4$-$3) emitting region at a location very
close to the quasar optical coordinates (offset by $\delta\rm{RA}=0\farcs 6$,
$\delta \rm{Dec}=0\farcs 5$). The CO(4$-$3) line profile is displayed in Fig.\
2. A Gaussian fit provides the following parameters: FWHM of
$390\pm38$ km s$^{-1}$, peak frequency at $109.778\pm 0.005$ GHz 
(corresponding to a redshift of 3.200), 
and intensity of $5.96\pm0.45$ Jy$\cdot$\kms . 
The CO(9$-$8) transition was not detected in the 1mm window. A Gaussian fit with
fixed line width and central position analogous to the CO(4$-$3)
transition showed only continuum at the position. A local peak close
to 100 \kms in the object's velocity frame is below the $3\sigma$
level. By combining both sidebands of the 1mm receiver, we obtain
a continuum detection of
$6.75\pm 1.32$ mJy.

\begin{figure} 
\psfig{file=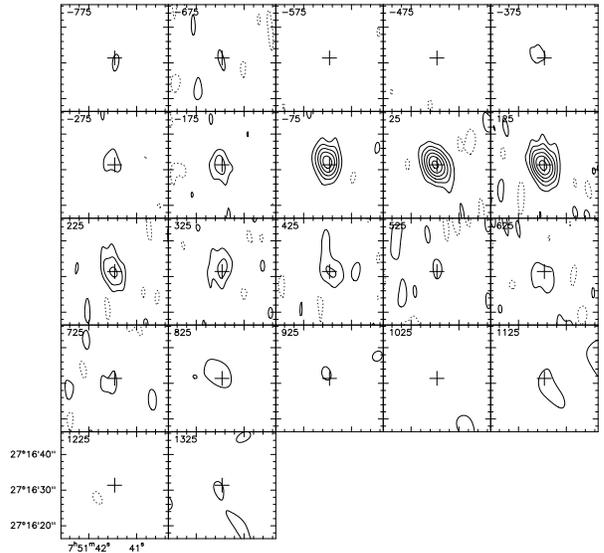,width=0.62\textwidth,angle=270} 
\caption{Channel maps of the CO($4-3$) line emission in MG0751+2716
with a contour spacing of 2.5 mJy/beam (the  
zero contour is not shown). The
cross indicates the optical position of the quasar.
Two different spectral setups were 
merged which overlap in the [$-75$,725] km s$^{-1}$ channels, 
resulting in three 
different rms levels over the spectral range (see Fig.\ 2 error bars). 
For the deconvolution of each channel, the appropriate combined UV 
coverage was taken into account.
The continuum flux has been subtracted.}
\label{fig-image} 
\end{figure} 

\begin{figure}
\psfig{file=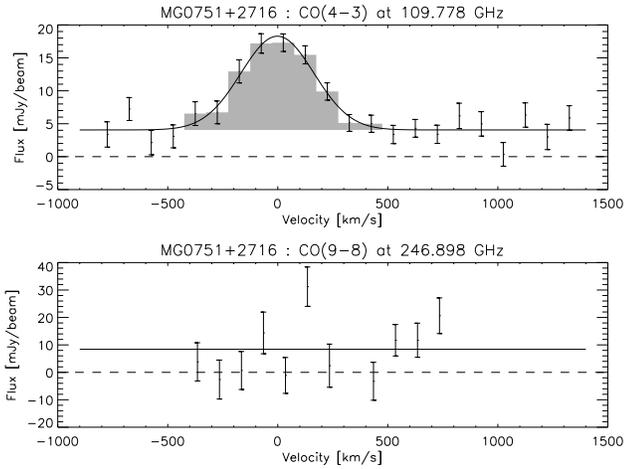,width=0.48\textwidth,angle=0}
\label{fig-image}
\caption{Spectra in CO($4-3$) ({\it upper panel}) and CO($9-8$)
({\it lower panel}) for  MG0751+2716, binned into 100~\kms\ wide 
channels.}
\end{figure}

In order to compute the CO line luminosity and total gas mass, a lensing correction 
must be applied.  
In spite of several efforts to model the lens system towards MG0751+2716 
(Leh\'ar et al.\ 1997; Tonry \& Kochanek 1999), 
additional work remains to be done: the lens appears to be a 
quite complex system which requires more shear than accounted 
for by the lensing galaxy (identified as G3 in Leh\'ar et al.\ 1997). 
This extra shear might come from what appears to be a group or cluster 
of galaxies, indicated by the large number of galaxies in the field 
around the quasar. Very recent modeling 
by J. Leh\'ar and B. McLeod (2001, private communication), 
based on HST optical imaging, provides 
an estimated optical magnification of 16.6. While differences 
between optical and CO/submillimeter net magnifications can be expected 
in some cases because of the difference in source sizes (subpc 
versus tens to hundreds of pc), models suggest that for optical 
magnifications less than about 20 the differences are generally not 
expected to be large (see Fig.\ 1 of Barvainis and Ivison 2002, and associated 
discussion). Therefore, we correct the CO emission for a magnification 
of factor of 16.6 and derive a line
luminosity $L'_{\rm CO} = 3.9\times 10^9$~K$\cdot$kms$^{-1}$~pc$^2$ ($H_0 = 75$~km
s$^{-1}$ Mpc$^{-1}$, $q_0 = 0.5$).

The molecular material emitting in the CO(4$-$3) transition is most 
likely close to the quasar, as in the case of the CloverLeaf (Kneib et 
al.\ 1998) and APM 08279+5255 (Downes et al.\ 1999),  and it is also
probably dense 
and warm. In order to calculate the total molecular gas mass we have 
considered two values for the ratio $M_{\rm H_2}/L'_{\rm CO}$: 
$0.8 M_{\sun}$~(K$\cdot$kms$^{-1}$~pc$^2$)$^{-1}$,
following Downes \& Solomon (1998) for nuclear rings in ultraluminous galaxies,
and $0.4 M_{\sun}$~(K$\cdot$kms$^{-1}$~pc$^2$)$^{-1}$ following Barvainis et al.\
(1997) for the molecular ``torus" 
in the Cloverleaf.  These conversion factors yield 
a total molecular gas mass in the range 
$M_{\rm H_2} = 1.6-3.1\times 10^9 M_{\sun}$.

It is of interest as well to compare the gas mass to the dynamical 
mass derived from the observed CO line width. With a derived Keplerian 
velocity of 400/sin($i$) km s$^{-1}$, and assuming that the molecular gas is 
located at a radius of 200 pc (see above references for the CloverLeaf 
and APM 08279+5255), we obtain a lower limit of $ M_{\rm dyn} > 
1.8\times 10^9 M_{\sun}$,
consistent with the derived value of $M_{\rm H_2}$.

\section{Concluding remarks and future prospects}

Though the present survey yielded a low detection rate in CO, there are several
new lensed quasar candidates yet to be observed based on their 
strong 850~$\mu$m continua, recently discovered in the course of the 
submillimeter 
survey by Barvainis \& Ivison (2002).  Supplementary, expanded-frequency
observations of some sources (most notably RXJ0911+055) may turn up 
more CO detections from the present source list.  

As for MG0751+2716, the centimeter radio source has four components connected 
by arcs (Leh\'ar et al.\ 1997), and in the optical it appears as 
a $1\arcsec$ diameter 
Einstein Ring.  A primary driver for this project was to find lensed
sources that could be spatially resolved in 
CO line emission.  This is currently possible for MG0751+2716 using the 
PdBI, and, like the Cloverleaf, reconstruction of the molecular
source structure and 
kinematics on very small angular scales using the lensing 
properties may prove to be quite interesting.

\acknowledgements{ We warmly thank all the IRAM staff who performed in 
service mode, with the Plateau de Bure Interferometer, all of the observations 
related to this project.  We also thank Ski Antonucci for important early 
contributions.  
Joseph Leh\'ar and Brian McLeod generously provided an estimate of the magnification
of MG0751+2716 in advance of publication.  DA and RB wish to thank the 
IRAM institute in Grenoble for hospitality during visits there. The NASA Extragalactic
Database (NED) and the CASTLeS compilation of lenses
(http://cfa-www.harvard.edu/castles) were used extensively in the course
of this work.
}

\end{document}